\documentclass[prb,twocolumn,,floatfix]{revtex4}
\usepackage{graphicx}
\usepackage{bm}
\usepackage{amssymb}
\usepackage{color}

\newcommand{\be}{\begin{equation}}
\newcommand{\ee}{\end{equation}}

\def \be{\begin{equation}}
\def \ee{\end{equation}}
\def \ba{\begin{array}}
\def \ea{\end{array}}
\def \bea{\begin{eqnarray}}
\def \eea{\end{eqnarray}}

\begin{document}

\title{Superconductivity in zigzag CuO chains}
\author{Erez Berg$^{1,3}$}
\email{berez@stanford.edu}
\author{Theodore H. Geballe$^2$}
\author{Steven A. Kivelson$^1$}
\affiliation{$^1$ Department of Physics, Stanford University, Stanford CA 94305-4045, USA
\\
$^2$ Department of Applied Physics, Stanford University, Stanford,
California 94305-4045, USA \\
$^3$ Department of Condensed Matter Physics, Weizmann Institute of Science,
Rehovot, 76100, Israel}

\begin{abstract}
Superconductivity has recently been discovered in Pr$_{2}$Ba$_{4}$Cu$_{7}$O$%
_{15-\delta }$ with a maximum $T_c$ of about 15K. Since the CuO planes in
this material are believed to be insulating, it has been proposed that the
superconductivity occurs in the double (or zigzag) CuO chain layer.
On phenomenological grounds
we propose a theoretical interpretation of the
experimental results in terms of a new phase for the zigzag chain, labelled
by C$_1$S$_\frac{3}{2}$. This phase has a
gap in the relative charge mode and a partial gap in the relative
spin mode.
It has gapless uniform charge and spin excitations and can have a
divergent superconducting susceptibility, even for repulsive
interactions. A microscopic model for the zigzag CuO chain is
proposed, and on the basis of density matrix renormalization group
(DMRG) and bosonization studies,
we adduce evidence that supports our proposal.
\end{abstract}

\date{\today}
\maketitle

\emph{Introduction - }The discovery of high-$T_{c}$ superconductivity has
raised the possibility of new mechanisms of superconductivity in strongly
correlated electron systems, which are radically different from the well
established BCS-Eliashberg mechanism. However, the lack of theoretical
techniques for obtaining well controlled solutions of even the simplest
models of strongly correlated fermions in dimensions greater than one has
seriously limited the understanding of such mechanisms.

An important exception is one dimensional systems, where powerful analytical
and numerical techniques are available. A one dimensional quantum system
cannot have superconducting long-range order even at zero temperature, as a
consequence of the Mermin-Wagner theorem. Nevertheless, such a system can
have a superconducting susceptibility which diverges as $T\rightarrow 0$.
Therefore, an array of coupled one dimensional systems can become a ``bulk"
superconductor with a reasonably high $T_c$ even in the limit where the
inter-chain coupling is very weak. This mechanism of superconductivity has
been studied extensively \cite{Q1DSC}. It serves as one of the only
well-established proofs of principle for superconductivity in a model with
purely repulsive interactions.

Among the candidates for an experimental realization for this mechanism are
the organic conductors and the ladder compound Sr$_{14-x}$Ca$_x$Cu$_{24}$O$%
_{41}$. Recently, superconductivity was discovered in Pr$_{2}$Ba$_{4}$Cu$%
_{7} $O$_{15-\delta }$ (Pr-247) in a certain region of oxygen reduction $%
\delta $, with a maximum $T_c$ of about 15K\cite%
{YamadaSC,YamadaSC1,YamadaNQR}. This material is isostructural with Y$_{2}$Ba%
$_{4}$Cu$_{7}$O$_{15-\delta }$ (YBCO-247). However, the two materials
display a dramatic difference in their electronic behavior: the copper-oxide
planes in YBCO\ are conducting, and are believed to play a crucial role in
the high-$T_{c}$ superconductivity in this material. In Pr-247 (as well as
in the closely related material Pr-248) the copper-oxide planes remain
antiferromagnetic with large moments even upon doping. To the best of our
knowledge, the conductivity of a single crystal of Pr-247 has not been
measured yet. In Pr-248, the conductivity 
is strongly anisortopic in the $a-b$ plane (with $\sigma _{a}/\sigma _{b}$
of up to 1000, where $a$ is the chain direction\cite%
{ResAnisotropy1,ResAnisotropy2,ResAnisotropy3}). This strongly suggests that
the copper-oxide planes in these materials are insulating, and that most of
the conductivity occurs in the metallic copper-oxide chain layers. Since the
double (or zigzag) chains are much more structurally robust than the single
chains, they are much less disordered and therefore are expected to be
better conductors.

Assuming that the electrical conductivity in Pr-247 (as in Pr-248) comes
mostly from the double chains 
it follows that the superconductivity 
must originate in these chains as well. 
While this assumption remains to be tested, existing NQR\ experiments\cite%
{SasakiNQRarchive} which measure the site-resolved spin-lattice relaxation
rate ($1/T_{1}$) have shown that only the double chain copper nuclear spins
show any sharp feature (a cusp) in their $1/T_{1}(T)$ curves at the
superconducting transition temperature, supporting the idea that the
superconductivity is intimately related to the double chains. This
possibility can have important implications in other materials which also
share the same zigzag CuO chain structure, such as YBCO-247, YBCO-248 and
the ladder compounds.

The purpose of this paper is to study the possibility of superconductivity
in zigzag CuO chains. This problem has been addressed in several previous
papers\cite{YamadaTheory,DMRG_zigzag_ladder,Nakano_fluc_exchange}. In Ref. [%
\onlinecite{YamadaTheory}], the CuO zigzag chain was treated in the weak
coupling limit and in the Hartree-Fock approximation. Ref. [%
\onlinecite{DMRG_zigzag_ladder}] studied a zigzag Hubbard ladder using DMRG.
In Ref. [\onlinecite{Nakano_fluc_exchange}] weak coupling fluctuation
exchange (FLEX) theory was used to study superconductivity in the zigzag
chains.

In this paper, we start from the experimental data, and analyze
the experimental constraints on the zigzag chain superconductivity
scenario. Then, we present a microscopic model for a single zigzag
chain, which contains in our view an important piece of physics
that has been omitted in previous studies, namely the oxygen $p_y$
orbitals and the ferromagnetic Hund's rule coupling on the oxygen
sites.\footnote{These effects have, however, been
considered in the context of the zigzag chains in the ladder materials\cite%
{RiceLadders}.} The model is studied using numerical density
matrix renormalization group (DMRG) calculations, as well as
analytic
renormalization group (RG) and 
strong coupling methods. 
The solution of the model shows features that are in agreement with 
experiment, namely the absence (or 
extremely small magnitude) of the spin gap and a
tendency towards superconductivity.

\section{Possible phases of the zigzag chain}

Our main assumption is that the zigzag CuO chains in Pr-247 are weakly
coupled, so the basic properties of the system at $T>T_c$ can be understood
in terms of the properties of 
decoupled chains. Let us review briefly a mechanism of superconductivity in
weakly coupled 1d systems. A 1d system typically has several types of
fluctuating order which coexist with each other (for example,
superconducting and spin or charge density wave orders). Even though none of
these can truly become long-range ordered in the isolated 1d system at
generic (incommensurate) filling, the susceptibility of the system to these
types for order can become large at low temperature. Then, treating the
inter-chain coupling at the mean-field level, the critical temperature is
determined by the Stoner criterion: $J\chi (T_c)=1$, where $J$ is the
inter-chain coupling and $\chi (T)$ is the susceptibility to the type of
order considered. Assuming that the inter-chain coupling constants of the
various types of order are all of comparable size, the type of order that is
most likely to 
be selected is the one which has the largest (i.e. most divergent)
susceptibility at low temperature.

A single-component 1d electron system with repulsive interactions can
generically be described at low energies as a Luttinger liquid with one
gapless spin and one gapless charge mode. Such a system is usually a poor
superconductor, with a superconducting susceptibility $\chi_{SC} \sim
T^{1/K_c-1}$, where $K_c$ is the charge Luttinger parameter. Since typically
$K_c<1$, the superconducting susceptibility is non-divergent while the CDW
and SDW susceptibilities both diverge.

Multi-component 1d systems are more promising in this respect. The problem
of two coupled Luttinger liquids has been studied extensively, and it is
found to be in a ``Luther-Emery" phase over a wide range of parameters. In
this phase, only the total charge mode remains gapless while all other spin
and charge modes are gapped. It is therefore similar to the phase of a
single component system with a spin gap caused by attractive interactions,
although the identity of the physical correlation functions in the two
systems is somewhat different. If we label the 1d phases as C$_n$S$_m$,
where $n$ and $m$ are the numbers of gapless charge and spin modes
respectively\cite{BalentsFisher}, the Luther-Emery phase is C$_1$S$_0$.
Since the entire spin sector is gapped, the dominant fluctuations in this
system are superconducting fluctuations and $4k_F$ CDW fluctuations, with
susceptibilities $\chi_{SC} \sim T^{1/2K_{+,c}-2}$ and $\chi_{CDW,4k_F} \sim
T^{2K_{+,c}-2}$ respectively. (Here, $K_{+,c}$ is the total charge mode
Luttinger parameter.) Near half filling, it has been shown \cite%
{SchultzLadderLowDope} that $K_{+,c} \sim 1$ and therefore superconducting
fluctuations always dominate. The interpretation of superconductivity in
Pr-247 as deriving from a Luther-Emery phase was proposed in Ref. [%
\onlinecite{YamadaTheory}].

An important observation is that in this scenario, the spin gap $\Delta_s$
must be significantly 
larger than $T_c$. At temperatures higher than $%
\Delta_s$, the spin modes are essentially gapless, and the exponents
controlling the susceptibilities would cross over to new exponents of a
gapless phase. In the regime $T_c<T< \Delta_s$ ``pseudogap" behavior should
be observed: a gap appears in the spectrum, but this gap is not associated
with any type of long-range order.

However, NQR measurements of the spin-lattice relaxation rate $1/T_1$ above $%
T_c$ show no evidence of 
gapped (i.e. thermally activated) behavior\cite%
{YamadaNQR,SasakiNQRarchive}. Assuming that the relaxation process is mostly
due to local magnetic field fluctuations (rather than electric field
gradient fluctuations), this implies that there is no spin gap in the zigzag
chains. The $1/T_1$ signal seems to follow a non-trivial power law as a
function of temperature, which implies a power law decay of spin-spin
correlations. For a discussion of the interpretation of $1/T_1$ NMR
measurements, see [\onlinecite{ScalapinoNMR}].

The preceding analysis leads us to the phenomenological idea that the zigzag
chain is in a phase with gapless spin excitations. Two uncoupled chains are
in a C$_2$S$_2$ phase. Inter-chain interactions can gap some of these modes.
For example, we can consider a C$_1$S$_2$ or a C$_1$S$_1$ phase. In these
cases, the spin-spin correlations behave as a power law (since the total
spin mode is gapless). However, some of the relative spin and charge modes
(the modes that correspond to fluctuations of the charge of the two chains
relative to each other) are gapped, which can enhance superconducting (as
well as other) susceptibilities.
In the rest of this paper, we will advocate an explanation of
superconductivity in Pr-247 in terms of such a phase, whose existence will
be established in a microscopic model for the zigzag chain.

The most likely candidate phase that emerges from our analysis is what we
will call a C$_1$S$_\frac{3}{2}$ phase, in which the relative charge mode
and ``half" of the relative spin mode are gapped. (We give a more precise
definition for these terms in Appendix \ref{app:c_1s_1.5}.) The resulting
susceptibilities in this phase are
\begin{eqnarray}
\chi_{SC} &\sim& T^{1/2K_{+,c}-\frac{5}{4}} \\
\chi_{CDW,4k_F} &\sim& T^{2K_{+,c}-2} \\
\chi_{CDW,2k_F} &\sim& \chi_{SDW,2k_F} \sim T^{K_{+,c}/2-\frac{5}{4}}
\end{eqnarray}
Note that the superconducting susceptibility is divergent for $K_{+,c}>0.4$,
and is the dominant one when $K_{+,c}>1$.

We comment that the proposed C$_1$S$_\frac{3}{2}$ phase is likely not to be
a true $T=0$ phase: it will be ultimately unstable to formation of a C$_1$S$%
_0$ phase. Nevertheless, if the gap in the total spin sector is small
compared to $T_c$, the physics above $T_c$ is best described in terms of a C$%
_1$S$_\frac{3}{2} $. Such a hierarchy of gaps is found to emerges in the
microscopic model that will be studied here (see Section \ref{sec:DMRG}).

Finally, the Luttinger exponent $K_{+,c}$ that controls the low-energy
properties of the system depends on microscopic details, and is therefore
difficult to estimate theoretically. However, since the power with which the
spin susceptibility depends on temperature also depends on $K_{+,c}$, it can
be extracted from a measurement of this susceptibility. From the NQR
measurement of $1/T_1T \sim \chi_{SDW}$ we extract the value $K_{+,c}
\approx 3/2$. 
Note that this corresponds to a regime of dominant
superconducting fluctuations with $\chi_{SC} \sim T^{-11/12}$.

\section{The model}

\begin{figure}[t]
\centering
\includegraphics[width=8cm]{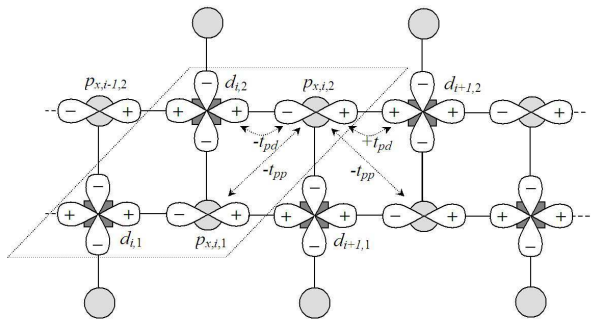}
\caption{The geometry of the zigzag CuO chain. Squares mark copper
atoms and circles are oxygen atoms. Also shown are a copper
$d_{x^2-y^2}$ orbitals and an oxygen $p_x$ orbitals. The dotted
box marks the unit cell.} \label{fig:zigzag}
\end{figure}

We now present a model for a single zigzag CuO chain. The geometry of the
zigzag chain is shown in Fig. \ref{fig:zigzag}. Of this structure, we will
ignore the outer oxygens, leaving two copper and two oxygen atoms per unit
cell. 
The relevant orbitals are the $3d_{x^{2}-y^{2}}$ on the Cu sites, and the $%
2p_{x}$ and $2p_{y}$ orbitals on the O sites. We assume that the on-site
Coulomb repulsion is large on both Cu and O sites, so that doubly occupied
states can be projected out, leaving an effective magnetic exchange
interaction. The Hamiltonian for holes can then be written as
\begin{equation}
H=H_{0}+H_{ex}  \label{Hmodel}
\end{equation}%
Here $H_{0}$ is the hopping Hamiltonian and $H_{ex}$ is the
magnetic spin exchange Hamiltonian. Let us first consider the
terms in $H_{0}$. A $3d$ Cu orbital can hybridize strongly with
the $p_{x}$ orbital of its neighbor O in the same chain, or a
$p_{y}$ orbital of the neighboring oxygen in the opposite chain.
However, it cannot hybridize with the nearest $p_{x}$ orbital in
the opposite chain or a $p_{y}$ orbital in the same chain due to
symmetry. Further neighbor hopping matrix elements (such as a
direct oxygen-oxygen hopping $t_{pp}$ shown in Fig.
\ref{fig:zigzag}) will be neglected, except in section
\ref{sec:tpp} where their effect on the main results will be
examined. For simplicity, we also also project out the $p_{y}$
orbitals, since they essentially ``belong" to the neighbor Cu $d$
orbital, in the sense that $H_{0}$ only connects a $p_{y}$ to that
$d$ orbital. The effect
of the $p_{y}$ orbitals will be reintroduced as an effective interaction in $%
H_{ex}$. Taking these considerations into account, $H_{0}$ is simply
\begin{eqnarray}
H_{0} &=&t_{pd}\sum_{i,\alpha ,\sigma }\hat{P}\left( d_{i,\sigma
,\alpha }^{\dagger }p_{x,i+1,\sigma ,\alpha }-p_{x,i,\sigma
,\alpha }^{\dagger
}d_{i,\sigma ,\alpha }+h.c.\right) \hat{P}  \nonumber \\
&+&\varepsilon \sum_{i,\alpha ,\sigma }p_{x,i,\sigma ,\alpha }^{\dagger
}p_{x,i,\sigma ,\alpha }  \label{H_0}
\end{eqnarray}%
Here $d_{i,\sigma ,\alpha }$ and $p_{x,i,\sigma ,\alpha }$ are hole
annihilation operators in $3d$ Cu and $2p_{x}$ O orbitals respectively, $%
\hat{P}$ is a projection operator that imposes the no-double occupancy
constraint, $i$ is the unit cell index, $\sigma =\uparrow ,\downarrow $ is
the spin index and $\alpha =1,2$ is the chain index. The oxygen sites have
an on-site potential $\varepsilon >0$.

Next, we consider $H_{ex}$. This has the form
\begin{eqnarray}
H_{ex} &=&J_{1}\sum_{i,\alpha }\left( \mathbf{S}_{d,i,\alpha }\mathbf{\cdot S%
}_{p_{x},i,\alpha }-\frac{1}{4}n_{d,i,\alpha }n_{p_{x},i,\alpha }\right.
\nonumber \\
&+&\left. \mathbf{S}_{p_{x},i,\alpha }\cdot \mathbf{S}_{d,i+1,\alpha }-\frac{%
1}{4}n_{p_{x},i,\alpha }n_{d,i+1,\alpha }\right)  \nonumber \\
&-&J_{2}\sum_{i}\left( \mathbf{S}_{d,i,1}\mathbf{\cdot S}_{p_{x},i,2}-\frac{1%
}{4}n_{d,i,1}n_{p_{x},i,2}\right)  \label{H_ex}
\end{eqnarray}%
Here $\mathbf{S}_{d}=\sum_{\sigma ,\sigma ^{\prime }}d_{\sigma }^{\dagger }%
\vec{\sigma}_{\sigma \sigma ^{\prime }}d_{\sigma ^{\prime }}$ where $\vec{%
\sigma}=\left( \sigma ^{x},\sigma ^{y},\sigma ^{z}\right) $ are Pauli
matrices, $n_{d}=\sum_{\sigma }d_{\sigma }^{\dagger }d_{\sigma }$ and
similar definitions for $\mathbf{S}_{p_{x}}$, $n_{p_{x}}$. $J_{1}>0$ is the
usual antiferromagnetic superexchange interaction. $J_{2}$, however, has a
qualitatively different physical origin: it comes from projecting out the
state with a doubly occupied oxygen site with one $p_{x}$ state and one $%
p_{y}$ state occupied by holes. Since the two holes belong to different
orbitals, they are subject to a \emph{ferromagnetic} Hund's rule interaction%
\cite{RiceLadders}. Therefore the effective exchange interaction is
ferromagnetic $(J_{2}>0)$ in this case. (Note the minus sign in Eq. \ref%
{H_ex}.)

The model we have derived for the zigzag chain has several new features that
distinguish it from the well-studied Hubbard ladder model. Firstly, as we
have seen, the inter-chain hopping is small relative to the intra-chain
bandwidth. (It arises only from further neighbor hoppings, such as the O-O
hopping
.) The coupling between the two chains thus comes mostly from the
electron-electron interactions\cite{SheltonLadders}. Secondly, the
inter-chain effective exchange interaction is ferromagnetic.

\section{Half filling}

\label{sec:half_filling}

It is instructive to consider the case of half filling in the double chain,
which we consider as the parent insulating state. In 
this case (i.e. one hole per Cu site), there is a charge gap of order ${%
\varepsilon }$. Then the problem essentially reduces to that of the zigzag
Heisenberg chain, described by the effective Hamiltonian

\begin{eqnarray}
H_{eff} &=&\tilde{J}_{1}\sum_{i,\alpha }\mathbf{S}_{d,i,\alpha }\mathbf{%
\cdot S}_{d,i+1,\alpha }  \nonumber \\
&&-\tilde{J}_{2}\sum_{i}\left( \mathbf{S}_{d,i,1}\mathbf{\cdot S}_{d,i,2}+%
\mathbf{S}_{d,i,1}\mathbf{\cdot S}_{d,i+1,2}\right)  \label{Heis}
\end{eqnarray}

Note that $\tilde{J}_{1},\tilde{J}_{2}$ are different from $J_{1},J_{2}$ in (%
\ref{H_ex}) $-$ they arise from projecting out the $p_{x}$ orbitals.
However, we still find that $\tilde{J}_{1}>0$, $\tilde{J}_{2}>0$. The model (%
\ref{Heis}) was studied in Ref. [\onlinecite{WhiteZigzag}] by a combination
of DMRG and bosonization, for $\tilde{J}_{2}$ both positive and negative. It
was found that for $\tilde{J}_{2}>0$, the coupling between the two chains is
irrelevant. This is due to the fact that the interaction between the two
chains in Eq. (\ref{Heis}) is geometrically frustrated. Later studies found
that the interaction actually contains a marginally relevant operator\cite%
{ItoiZigzag,NersesyanZigzag}, but the RG flow is extremely slow, so the
system can still be regarded as gapless for all practical purpose\cite%
{ItoiZigzag}. This is an important 
difference between the zigzag ladder and the simple Hubbard ladder; the
latter has a large spin gap $\Delta_s\sim J/2$ at half filling.

\section{DMRG\ simulations}

\subsection{$t_{pp}=0$ case}

\label{sec:DMRG}

\begin{figure}[t]
\centering
\includegraphics[width=8cm]{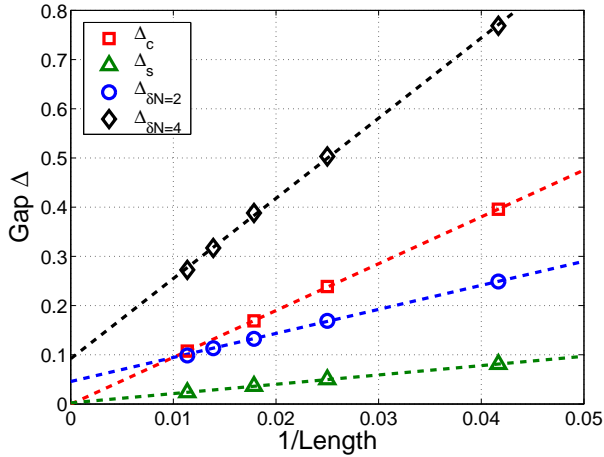}
\caption{Calculated gaps from DMRG, measured in units of $t_{pd}$, as a
function of $1/L$, where $L$ is the number of Cu sites in the system. $%
\Delta_c$ is the charge gap, $\Delta_s$ is the spin gap, and $\Delta_{%
\protect\delta_N=2}$ and $\Delta_{\protect\delta_N=4}$ are the gaps for
transferring one and two holes from one chain to the other (see text). The
dashed lines are linear extrapolations.}
\label{fig:DMRG_gaps}
\end{figure}

\begin{figure*}[tbp]
\centering
\includegraphics[width=10cm]{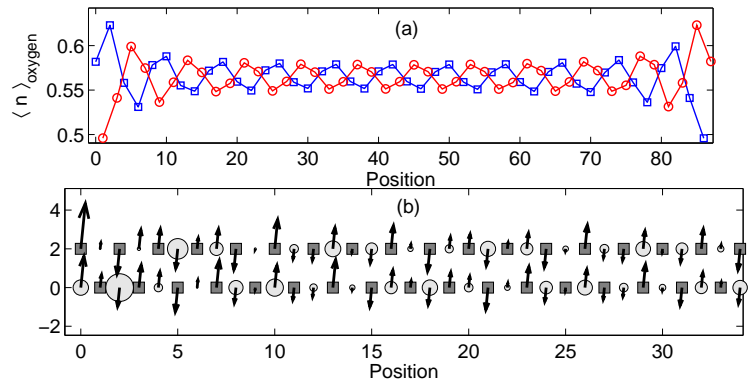}
\caption{(a) The average hole density on oxygen sites as a function of
position. Blue Squares (red circles) mark the upper (lower) chain from the
DMRG simulation with $\langle n \rangle=1.25$ holes per Cu site. (b) $%
\langle S^z_i \rangle$ as a function of position. Copper sites are marked by
squares, and oxygen sites - by circles. The size of the circles represents
the average hole density on the oxygen sites (bigger circles mean larger
hole density).}
\label{fig:DMRG_profiles}
\end{figure*}

The doped zigzag chain cannot be understood as simply as the half filled
one. 
Moreover, it is not entirely clear what doping level leads to
superconductivity in Pr-247; counting charges (and assuming that the valence
of the Pr ions is $+3$), there are $9/7- 2\delta/7$ doped holes per copper
site on average, and superconductivity occurs when $\delta \gtrsim 0.3$ (see
[\onlinecite{YamadaSC1}]). However, these holes are distributed in an
unknown manner between the plane, single chain, and double chain Cu sites.
Therefore, we have performed DMRG simulations with a finite doping
concentration which we take, somewhat arbitrarily, to be 0.25 ``doped
holes'' per Cu (\textit{i.e.} the density of holes is taken to be $n=1.25$
per Cu site). The following parameters were used in Eq. (\ref{Hmodel}): $%
t_{pd}=1$, $\varepsilon=0.5$, and $J_1=J_2=0.5$ 
The values for $t_{pd}, J_1$, and $\varepsilon$ are 
comparable to values that are commonly used in effective $t-J$ models for
the copper-oxide planes in the cuprates. In the simulations described in
this section, the direct oxygen-oxygen hopping, $t_{p,p}$, was neglected. 
(The effect of non-zero $t_{p,p}$
is examined in the next subsection.)
Up to $m=2700$ states per block where kept for the longest systems ($L=88$
Cu sites), resulting in an average truncation error of about $10^{-6}$. The
energies where extrapolated to zero truncation errors using the standard
procedure\cite{WhiteExtrap}. The convergence of the ground state energy as a
function of the number of sweeps improves dramatically when a small
perpendicular copper - oxygen term is added to the hopping Hamiltonian (\ref%
{H_0}), with $t_{\perp}=0.01$. We have checked that the physical results are
not affected by adding this term.

In Fig. \ref{fig:DMRG_gaps} we plot various gaps in the spectrum as a
function of $1/L$. $\Delta_c=E_{N+2}+E_{N-2}-2E_{N}$ is the charge gap, and $%
\Delta_s=E_{S^z=2}-E_{S^z=1}$ is the spin gap. We define the spin gap
relative to the $S^z=1$ state, since the ground state is found to have spin
one for the systems considered here\footnote{%
It turns out that the spin of the ground state is determined by the parity
of the number of holes in each chain: if the number is odd, the ground state
has spin 1 (which was the case in all the simulations presented here), and
if it is even the ground state has spin 0. This is similar to the situation
in the spin-1 Heisenberg chain with open boundary conditions: the ground
state is a triplet for an odd length chain and a singlet for an even length
chain. In the spin-1 chain, however, the first excitation is localized at
the edges of the chain and its separation from the ground state decays
exponentially with the length.}. In addition, we have calculated the gap to
the state with one or two holes transferred from one chain to the other: $%
\Delta_{\delta N=2}=E_{\delta N=2}-E_{\delta N=0}$ and $\Delta_{\delta
N=4}=E_{\delta N=4}-E_{\delta N=0}$, where $\delta N=N_2-N_1$ is the
difference between the number of holes in the two chains. The direct
measurement of $\Delta_{\delta N}$ is possible due to the fact that $N_1$
and $N_2$ (the numbers of holes on each chain) are separately conserved. In
the presence of a small inter-chain hopping term, this conservation law is
weakly violated. However, it is still possible to measure $\Delta_{\delta N}$
by applying an inter-chain potential difference $\Delta V$. When $\Delta
V=\Delta_{\delta N}$, one hole is transferred from one chain to the other.
This transition is smoothed by the inter-chain hopping term, with a width
proportional to $t_{\perp}$. For $t_{\perp}=0.01$, we found that this does
not considerably limit the accuracy in the determination of $\Delta_{\delta
N}$.

The gaps were extrapolated linearly with $1/L$ to the thermodynamic limit ($%
L\rightarrow \infty$). The charge gap $\Delta _{c}$ extrapolates to very
close to zero. The spin gap $\Delta _{s}$ also extrapolates to a very small
value, which is indistinguishable from zero to the accuracy of our
calculations. However, $\Delta _{\delta N=2}$ and $\Delta _{\delta N=4}$
clearly extrapolate to finite values. The linear extrapolation results are $%
\Delta _{\delta N=2}(1/L \rightarrow 0) \approx 0.046$, $\Delta _{\delta
N=4}(1/L \rightarrow 0) \approx 0.097$.

Let us assume that, as the simulation suggests, $\Delta _{c}=\Delta _{s}=0$.
In terms of the classification by the number of gapless bosonic modes, this
implies that we must have at least one gapless charge mode and one gapless
spin mode. However, since $\Delta_{\delta N} \ne 0$ at least some of the
relative spin and charge modes of the two chains are gapped by the
inter-chain coupling. The fact that $\Delta_{\delta N=4} \ne 0$ in the
thermodynamic limit implies that \emph{the relative charge mode is gapped}.
This is because it involves transferring two holes from one chain to the
other, and two holes can be combined to a singlet and therefore carry no
spin.

Fig. \ref{fig:DMRG_profiles}a shows the average hole density on the oxygen
sites in a chain of $L=88$ Cu sites. The density profile shows pronounced
density oscillations with a period of $\lambda _{c}=4a$ where $a$ is the
lattice constant. For the density of holes in the simulation, this
corresponds to a wavevector of $4k_{F}$. Fourier transforming the density
profile reveals that the $2k_{F}$ component is almost completely absent,
which is characteristic of systems with strong repulsion. Note also that the
relative
charge densities in the two chains appear
rigidly locked to each other so that the charge oscillations are strictly staggered,
further evidence of the existence of a relative charge gap.

A zigzag chain model similar to Eq. (\ref{Hmodel}) was studied in Ref. [%
\onlinecite{OgataRice}], and a spin gap was found. However, in that study
there where no oxygen sites and it was assumed that $J_{2}<0$ (i.e.
antiferromagnetic) and $|J_{2}|>J_{1}$. The ferromagnetic case was studied
(in a different context) in Ref. [\onlinecite{FMZigzag}], and their results
seem consistent with ours (at least over a certain range of parameters).

In Fig. \ref{fig:DMRG_profiles}b we show $\langle S_{i}^{z}\rangle $ along
with the oxygen site density (represented by the size of the circles) in the
same system. A weak Zeeman field $\mathbf{H}=0.01\mathbf{\hat{z}}$ was
applied to the first and last Cu sites on the chain in order to select a
direction in spin space. The main periodicity in the spin density is $%
\lambda _{s}=8a$, or $2k_{F}$. Note that a peak in the hole number density
is always accompanied by a $\pi $ phase shift in the spin density wave. Some
features of the pattern of the charge and spin density can be understood
from a strong coupling approach, as discussed in section V A.

\subsection{$t_{pp} \ne 0$ case}

\label{sec:tpp}

So far, we have neglected the oxygen-oxygen hopping $t_{pp}$. This term is
expected to be significantly smaller than $t_{pd}$, because of its longer
range. Estimates from LDA\ calculations \cite{LDA} give $t_{pp}\sim
0.27t_{pd}$\cite{YamadaTheory}.

As we saw, the DMRG simulation with $t_{pp}=0$ revealed a finite
gap to the relative charge mode between the two chains. In such a
phase, single particle inter-chain hopping is
irrelevant\cite{SheltonLadders}. Therefore, this phase is expected
to be robust over a finite range of $t_{pp}$. However, this
argument cannot tell us what is the range of stability. In order
to study the effect of a finite $t_{pp}$, we have performed
simulations including an oxygen-oxygen hopping term $H_{pp}$.
Using the same notation as in Eq. (\ref{H_0}), $H_{pp}$ is given
by

\begin{equation}
H_{pp}=-t_{pp}\sum_{i,\sigma }\hat{P}\left( p_{x,i,\sigma ,1}^{\dagger
}p_{x,i+1,\sigma ,2}+p_{x,i,\sigma ,2}^{\dagger }p_{x,i,\sigma
,1}+h.c.\right) \hat{P}  \label{Hpp}
\end{equation}

In the presence of this term, the method we used previously to determine
whether there is a relative charge gap (based on measuring the energy gap
for transferring one hole from one chain to the other) is no longer
applicable
because now the charges on the two chains are not
conserved separately.
Consequently, the charge difference cannot be controlled in the
calculation. Instead, we used an alternative method to determine in which
phase the system is. The local spin and charge densities on each chain where
measured and Fourier transformed. We define

\begin{eqnarray}
n_{q} &=&\frac{1}{\sqrt{L/2}}\sum_{i=1}^{L/2}e^{-iqx_{i}}n_{p_{x},i,1}
\nonumber \\
S_{q}^{z} &=&\frac{1}{\sqrt{L/2}}\sum_{i=1}^{L/2}e^{-iqx_{i}}S_{d,i,1}^{z}
\label{nq_Szq}
\end{eqnarray}

Here $n_{p_{x},i,1}=\sum_{\sigma }p_{x,i,\sigma ,1}^{\dagger }p_{x,i,\sigma
,1}$, $S_{d,i,1}^{z}=\frac{1}{2}\sum_{\sigma }\sigma d_{i,\sigma
,1}^{\dagger }d_{i,\sigma ,1}$ and $L$ is the length of the chain (number of
Cu\ sites). Due to
the open boundary conditions,
$n_{q}$ and $S_{q}^{z}$ show pronounced peaks at certain
wavevectors. (These are known as \textquotedblleft Friedel-like" oscillations%
\cite{WhiteFriedel}.) These peaks occur at the wavevectors of the
gapless spin/charge modes of the system. A 1D version of
Luttinger's theorem\cite{Affleck1dLuttinger} states that there
must be a gapless mode at a certain wavevector, corresponding in
our case to $4k_{F}$ of a single chain, where $2k_{F}=\pi n$ ($n$
is the number of holes per Cu site). In addition, there could be
other gapless modes at other wavevectors (e.g., charge or spin
modes at $2k_{F}$ for each chain, etc.). If there is only a single
gapless charge mode, as we found in the case $t_{pp}=0$, then we
expect to find gapless modes (and therefore peaks in the Fourier
transformed local spin/charge densities of a finite system with
open boundary conditions) only at a \emph{single} wavevector, plus
its harmonics.
 If, on the
other hand, the relative charge gap closes and there is more than
one gapless charge mode, there is no reason why these modes cannot
have different wavevectors (these are the analogues of the two
Fermi wavevectors in the non-interacting ladder problem). This
way, the phase transition from a phase with a single gapless
charge mode to a phase with two gapless modes can be identified.

\begin{figure}[t]
\centering
\includegraphics[width=9cm]{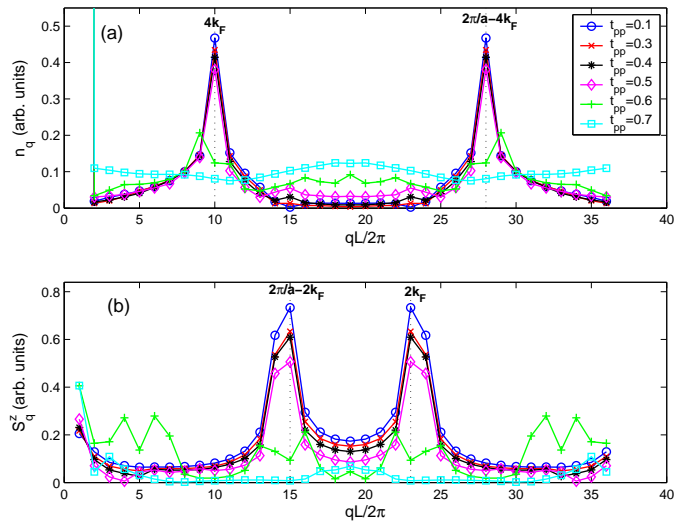}
\caption{(Color online.) Fourier transforms of the charge (a) and
spin (b) profiles (defined
in Eq. (\protect\ref{nq_Szq})) calculated by DMRG in a system of length $%
L=72 $ Cu sites. Various values of $t_{pp}$ (Eq.
(\protect\ref{Hpp})) were used. ($t_{pp}$ is measured in units of
$t_{pd}$.) Both profiles change
qualitatively and new peaks appear between $t_{pp}=0.5t_{pd}$ and $0.6t_{pd}$%
, which signals a phase transition to a phase with two independent
gapless charge modes.} \label{fig:nq_Szq}
\end{figure}

The results of the DMRG calculations are shown in Fig. \ref{fig:nq_Szq}. No
additional peaks appear at new wavevectors and the spin/charge profiles do
not change qualitatively until $t_{pp}\sim 0.5-0.6t_{pd}$. This shows that
the phase with a single gapless charge mode is robust at least up to $%
t_{pp}=0.5t_{pd}$, which is higher than the value estimated from band
structure calculations for the zigzag CuO chain.


\section{Analytical results in special limits}

\subsection{$t_{pd} \rightarrow 0$ limit}

\label{subsec:t0limit}

\begin{figure}[b]
\includegraphics[width=6cm]{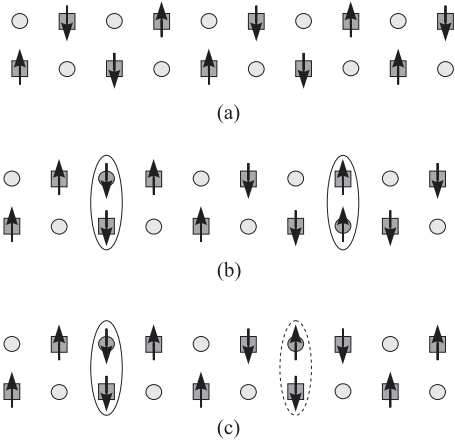}
\caption{Typical pattern of the spin correlations in various situations.
Copper (oxygen) sites are shown as square (circles). (a) The undoped chain.
The intra-chain interaction is antiferromagnetic, while the ferromagnetic
inter-chain interaction is frustrated. (b) Two doped holes, one in each
chain. Note that the phase of the spin fluctuation shifts by $\protect\pi $
around the doped holes. The spin of each doped holes is parallel to the spin
of the nearest copper site in the opposite chain, which gains some
ferromagnetic interchain exchange energy. (c) Two doped holes in the same
chain. This is similar to (b), except that now the inter-chain ferromagnetic
exchange energy of the hole on the right is \emph{lost}. Putting two
neighboring doped holes in the same chain is less favorable energetically.}
\label{fig:CuO_spins}
\end{figure}

We start from a qualitative strong coupling description, in which we assume
that 
$t_{pd}^2/\varepsilon\ll J_{1},J_{2}$. 
As we saw, at half filling the inter-chain coupling $J_{2}$ is frustrated,
so the low-energy spectrum is essentially like that of two decoupled chains.
A typical snapshot of the spin configuration in this state is depicted in
Fig. \ref{fig:CuO_spins}a.

We now consider a lightly hole doped system. Since all the copper sites are
occupied, the doped holes reside on 
oxygen sites. Fig. \ref{fig:CuO_spins}b shows two neighbor holes on opposite
chains. Since $J_1>0$, every doped hole induces a $\pi$ shift in the phase
of the spin density wave around it. The doped hole locally relieves the
frustration of the inter-chain ferromagnetic coupling $J_2$, causing a net
coupling between the spin fluctuations on the 
two chains. This coupling is proportional to the hole concentration $x$
times $J_2$. The hole can move along the chain via a second order process in
$t_{pd}$ without disturbing the local spin order. Note that some exchange
energy is gained also near the neighbor hole in the opposite chain.

However, if two neighbor holes are in the same chain (as shown in \ref%
{fig:CuO_spins}c), the situation is different. Now we cannot satisfy the $%
J_2 $ terms near both holes, and the inter-chain coupling is partially
frustrated. Therefore, 
the lowest energy states will be 
ones in which the holes appear \emph{in alternating order} in the two
chains. 
The holes are free to move so as to reduce their zero point kinetic energy,
as long as they do not pass each other.

The qualitative picture in the small $t_{pd}$ limit can explain some of the
features in Fig. \ref{fig:DMRG_profiles}b. As mentioned previously, the
peaks in the charge density on the O sites are accompanied by $\pi$ phase
shifts in the spin density wave on the Cu sites, which can be understood as
coming from the antiferromagnetic coupling between 
neighboring Cu and O spins. 
The fact that the hole density oscillations are $\pi$ phase shifted
between the two chains, 
is reminiscent of the alternating order between the chains described above.
Moreover, the spin on the O site with maximum local density is always \emph{%
parallel} to the spin of the nearest Cu on the opposite chain, due to the
ferromagnetic coupling between them. The DMRG picture is therefore
qualitatively very similar to Fig. \ref{fig:CuO_spins} even though $t_{pd}$
is not small.

In fact, we suspect that the present analysis is qualitatively correct for
small enough $x$ independent of the magnitude of $t_{pd}$. For a single hole
in an antiferromagnetic chain, a $\pi$ phase shift in the antiferromagnetic
correlations appears to be generic. Beyond that, the energy which favors
alternating holes on the two chains is purely geometric in origin, and so
only requires that the holes be sufficiently dilute.

\subsection{RG and Bosonization treatment}

Another limit of the zigzag chain model (\ref{Hmodel}) that can be treated
analytically is the weak inter-chain coupling limit ($|J_{2}|\ll J_{1},t$).
In this limit, only the low-energy degrees of freedom of the decoupled
chains are affected by the inter-chain coupling. These degrees of freedom
can be described as two Luttinger liquids. The inter-chain coupling can then
be treated using perturbative RG. The nature of the strong-coupling fixed
point can be understood using abelian bosonization. This is a standard
procedure\cite{BalentsFisher,CongjunFradkin}.

As we will see, in the weak inter-chain coupling limit a spin gap is
predicted, in contradiction to DMRG results of section \ref{sec:DMRG}. The
discrepancy can be explained by the fact that the inter-chain coupling is
not small. Rather, we will use bosonization to parameterize an effective low
energy model which is consistent with the numerical results. This model is
then used to calculate low-energy susceptibilities of the system.

The Hamiltonian is written as

\begin{equation}
H=H_{LL}+H_{int}  \label{H_boson}
\end{equation}
Here $H_{LL}$ is a Luttinger liquid Hamiltonian that describes the low
energy degrees of freedom of each chain, and $H_{int}$ is the inter-chain
coupling Hamiltonian, 

\begin{equation}
H_{LL}=H_{0,s}+H_{0,c}  \label{H_LL}
\end{equation}
where $H_{0,s}$ and $H_{0,c}$ are bosonic Hamiltonians for the spin and
charge modes of each chain:

\begin{equation}
H_{0,s}=\frac{u_{s}}{2}\sum_{i=1,2}\int dx\left[ K_{s}\left( \partial
_{x}\theta _{i,s}\right) ^{2}+\frac{1}{K_{s}}\left( \partial _{x}\varphi
_{i,s}\right) ^{2}\right]  \label{H0spin}
\end{equation}%
\[
H_{0,c}=\frac{u_{c}}{2}\sum_{i=1,2}\int dx\left[ K_{c}\left( \partial
_{x}\theta _{i,c}\right) ^{2}+\frac{1}{K_{c}}\left( \partial _{x}\varphi
_{i,c}\right) ^{2}\right]
\]
and $i=1,2$ is the chain index. SU$\left( 2\right) $ symmetry gives the
constraint $K_{s}=1$, while $K_{c}<1$ depends on the details of the
intra-chain microscopic interactions. Here $\phi$ and $\theta$ represent
dual fields, \textit{i.e.} $[\partial_x\theta_a(x),\phi_b(x^\prime)]= i
\delta_{a,b} \delta(x-x^\prime)$, and are related to the density and current
operators, \textit{e.g.} $\rho_{i}= \sqrt{2/\pi} \partial_x\phi_{i,c}$ and $%
j_{i}=u_c\sqrt{2/\pi}\partial_x\theta_{i,c}$ where $\rho_i$ and $j_i$ are,
respectively, the electron density and current density on chain $i$. At half
filling, the charge sector will also have a non-linear term $g_u \cos(\sqrt{%
8\pi} \varphi_{c,i})$, which corresponds to $4k_F$ umklapp scattering. The
spin sector 
should include marginally irrelevant operators, which we have not written
here.

Next, we consider the interaction term between the two chains. This term is
most conveniently written in terms of the corresponding fermionic degree of
freedom, $\psi _{i,\sigma }$, with $i=1,2$ for the two chains, which is in
turn written in terms of the right and left moving fermions, $\psi
_{i,\sigma }\sim e^{ik_{F}x}\psi _{i,R,\sigma }+e^{-ik_{F}x}\psi
_{i,L,\sigma }$, where $2k_{F}=\pi n$ is the Fermi wavevector ($n$ is the
total density of holes). Taking a naive continuum limit of the $J_{2}$ term
in (\ref{H_ex}), we get:
\begin{eqnarray}
H_{int} &=&J_{2}a\int dx{\LARGE \{}\ \mathbf{S}_{1}\left( x\right) \cdot
\mathbf{S}_{2}\left( x\right) +\mathbf{S}_{1}\left( x\right) \cdot \mathbf{S}%
_{2}\left( x+a\right)  \nonumber \\
&-&\frac{1}{4}\left[ n_{1}\left( x\right) n_{2}\left( x\right) +n_{1}\left(
x\right) n_{2}\left( x+a\right) \right] {\LARGE \}}  \label{H_int}
\end{eqnarray}%
Here $\mathbf{S}_{i}\left( x\right) =\psi _{i}^{\dagger }\left( x\right)
\vec{\sigma}\psi _{i}\left( x\right) $, $n_{i}\left( x\right) =\psi
_{i}^{\dagger }\left( x\right) \psi _{i}\left( x\right) $, and $a$ is the
lattice constant. This form satisfies the requirement that the Hamiltonian
is symmetric with respect to $\mathbf{S}_{1}\left( x\right) \rightarrow
\mathbf{S}_{2}\left( x\right) $, $\mathbf{S}_{1}\left( x\right) \rightarrow
\mathbf{S}_{2}\left( x+a\right) $ and similarly for the $n_{i}$'s, which
corresponds to a reflection that interchanges the two chains, followed by a
translation by one Cu-O distance.

In bosonized form, the most relevant part of (\ref{H_int}) is written as:
\begin{eqnarray}
H_{int} &=& -\frac{\sin(\frac{\delta a}{2})}{(\pi a)^2} \int dx \sin \left(
\sqrt{4 \pi} \varphi_{-,c}-\frac{\delta a}{2} \right)  \nonumber \\
&& \times \left[ g_{1} \hat {\mathcal{O}}_{s1}+g_{2} \hat {\mathcal{O}}_{s2} %
\right]  \label{H_int_bos}
\end{eqnarray}
Here $\hat {\mathcal{O}}_{s1} \equiv \left[ \cos \left( \sqrt{4\pi }\varphi
_{-,s}\right) +\cos \left( \sqrt{4\pi }\theta _{-,s}\right) \right] $ and $%
\hat {\mathcal{O}}_{s2} \equiv \left[ \cos \left( \sqrt{4\pi }\varphi
_{-,s}\right) +\cos \left( \sqrt{4\pi }\varphi _{+,s}\right) \right]$ where $%
\varphi _{\pm ,c}=\left( \varphi _{1,c}\pm \varphi _{2,c}\right) /\sqrt{2}$
are the even/odd charge modes of the two chains, and similarly for $\varphi
_{\pm ,s}$ and $\theta _{\pm ,s}$. We have defined $\delta =\pi(n-\frac{1}{a}%
)$, where $n$ is the density of holes per unit length. ($\delta $ is a
wavevector whose length is proportional to the amount of doping away from
half filling.) 
At half filling, additional umklapp terms appear. $g_{1}$ and $g_{2}$ are
dimensionless coupling constants, whose bare values at the initial scale
are: $g_{1}=-g_{2}=\frac{J_{2}a}{2}$. Under RG, $g_{1}$ and $g_{2}$ will
flow, and need not remain equal in magnitude. However, additional couplings
are prevented as long as the exact SU$(2)$ spin-rotational symmetry is
respected. Specifically, there are three distinct cosines of spin fields,
but only two independent coupling constants.

Note that the inter-chain coupling (\ref{H_int}) vanishes in the limit $%
\delta \rightarrow 0$, i.e. at half filling. This is due to the frustration
of the inter-chain coupling in that limit. (See section \ref%
{sec:half_filling}. ) Then marginal operators need to be considered. These
produce an extremely small gap in the spectrum, especially in the $J_2<0$
case \cite{ItoiZigzag}, so the system can be considered as essentially
gapless at half filling\cite{WhiteZigzag,ItoiZigzag}.

Away from half filling, both the $g_1$ and $g_2$ terms in
(\ref{H_int_bos}) are relevant, and produce a gap in the spectrum.
They both have the same scaling dimension of $1+K_{+,c}<2$
(assuming that $K_{\pm ,s}=1$). The system flows to a strong
coupling fixed point at which the $\varphi _{-,c}$, $\varphi
_{+,s}$ and $\theta _{-,s}$ fields are pinned. Then the only
gapless mode is the total charge mode, and the system is in a
C$_{1}$S$_{0}$
phase, similar to the generic situation in the Hubbard ladder\cite%
{BalentsFisher}.

However, the DMRG\ result indicates that while there is a substantial gap in
the relative charge 
mode, the gap in the total spin sector is either zero or very
small (see Fig. \ref{fig:DMRG_gaps}). This discrepancy is likely
to be caused by less relevant (sub-leading) operators, not
included in Eq. (\ref{H_int_bos}), whose bare coefficients are of
order unity (since we are not initially at weak coupling). The
neglect of these operators in the initial stages of the RG
transformation is 
a quantitatively unreliable approximation. Taking the DMRG\ result into
account, we \emph{hypothesize} that the effective value of $g_2$ \emph{at
the fixed point} is close to 0 (since non-zero $g_2$ is what induces a full
spin-gap). If this is true, then over a broad intermediate range of
energies, the system is governed by the unstable fixed point with $g_2=0$
and non-zero $g_1$.


We would like to stress that the smallness of the spin gap is not
a result of fine tuning, but rather appears (from the numerics) to
be a ubiquitous property of the model. (The same result is
obtained over a range of different parameters and doping levels.)
In contrast, if the inter-chain coupling $J_{2}$ is artificially
turned to be antiferromagnetic, a spin gap appears in the
numerical simulation. Therefore the small spin gap seems to be a
feature of the ferromagnetic $J_{2}$ case.

Note that the phase with $g_{1}\ne 0$, $g_{2}=0$ is unusual, because neither
the field $\varphi_{-,s}$ nor its 
dual $\theta_{-,s}$ can be pinned (since $\hat{\mathcal{O}}_{s1}$ contains
the cosine terms containing both fields 
with equal magnitude). The properties of this phase, which follow closely
from an earlier analysis of the 
Heisenberg ladder by Shelton, Nersessyan and Tsvelik\cite{SheltonTsvelik},
are explored in appendix \ref{app:c_1s_1.5}. The main conclusion is that
essentially ``half" of the relative spin mode is gapped, and the other half
is gapless. Therefore we denote this phase as C$_1 $S$_\frac{3}{2}$.

\subsection{Physical susceptibilities}

Given that the system is in a C$_1$S$_\frac{3}{2}$ phase, the physical
susceptibilities can be calculated in a similar way to that described in [%
\onlinecite{SheltonTsvelik}], with addition of the charge modes. For an
example of such a calculation, see appendix \ref{app:c_1s_1.5}. The leading
temperature dependence of these susceptibilities depends only on the total
charge Luttinger parameter $K_{+,c} $. The superconducting susceptibility is
\begin{equation}
\chi _{SC}\sim T^{\frac{1}{2K_{+,c}}-\frac{5}{4}}  \label{Sc_q}
\end{equation}
for both singlet and triplet pairing, which gives that the superconducting
susceptibility is divergent for $K_{+,c}>0.4$ (\textit{i.e.} even for
strongly repulsive interactions).

Other susceptibilities are $2k_{F}$ SDW and CDW, with
\begin{equation}
\chi _{SDW,2k_F}\sim \chi _{CDW,2k_F}\sim T^{\frac{K_{+,c}}{2}-\frac{5}{4}}
\end{equation}
and $4k_{F}$ CDW:
\begin{equation}
\chi _{CDW,4k_{F}}\sim T^{2K_{+,c}-2}  \label{4kf_CDW}
\end{equation}
The superconducting susceptibility is the most divergent one for $K_{+,c}>1$%
.

Assuming that the relative charge mode $\varphi_{-,c}$ is gapped, it is
possible to extract $K_{+,c}$ from the density profile in the DMRG
simulation (see Fig. \ref{fig:DMRG_profiles}). The density profile is
expected to behave as $\langle n_i \rangle \sim \cos(4k_F x_i)/[L\sin(\pi
x_i/L)]^{K_{+,c}}$, which is essentially the square root of the $4k_F$ part
of the density-density correlation function \cite{WhiteFriedel}. The
amplitude of the $4k_F$ CDW near the middle of the chain, $A_{x_i=L/2}$,
thus decays as $L^{-K_{+,c}}$. Fitting $A_{x_i=L/2}$ of chains with
different lengths to this expression, we obtain $K_{+,c}=0.6 \pm 0.05$.

Since NQR measures of the spin-lattice relaxation rate have been done on
superconducting Pr-247, it is interesting to extract the temperature
dependence of this quantity from the theory. Assuming that the main
relaxation mechanism is the coupling of the nuclear spins to fluctuations of
the local magnetic field due to electronic spins, $1/T_{1}$ is given by\cite%
{ScalapinoNMR}:

\begin{equation}
\frac{1}{T_{1}}=\frac{1}{2N}\sum_{q\alpha }\left\vert A_{\alpha }\left(
q\right) \right\vert ^{2}S^{\alpha \alpha }\left( q,\omega _{R}\right)
\label{T1NQR}
\end{equation}
where $A_{\alpha }\left( q\right) $ is the form factor of the hyperfine
Hamiltonian, $\omega _{R}$ is the nuclear spin resonance frequency, and $%
S^{\alpha \alpha }\left( q,\omega _{R}\right) $ is the electronic spin
structure factor. $\omega _{R}$ is typically smaller than any other energy
scale in the problem, so we will assume $\omega _{R}=0$.

The spin structure factor in the 
C$_1$S$_\frac{3}{2}$ is expected to be of the form:
\begin{equation}
S^{\alpha \alpha }\left( q,\omega =0\right) \sim A\ln q+Bq^{\frac{K_{+,c}}{2}%
-\frac{5}{4}}  \label{S_q_w}
\end{equation}
where $A$ and $B$ are constants. The two terms here are the two main
contributions to $S^{\alpha \alpha }(q,\omega=0)$ which come the vicinity of
the points $q=0,2k_{F}$. Integrating (\ref{S_q_w}) over $q$, we get the
dominant temperature dependence of the spin-lattice relaxation rate:
\begin{equation}
\frac{1}{T_{1}}\sim AT\ln T+BT^{\frac{K_{+,c}}{2}-\frac{1}{4}}
\label{T1Theory}
\end{equation}
NQR measurements\cite{YamadaNQR,SasakiNQRarchive} show that $T_{1}^{-1}$
behaves as a power law of temperature, with different exponents above and
below $T_{c}$. The fact that a power law is seen above $T_c$ is a clear
evidence for the absence of a spin gap, and is consistent with a C$_1$S$_{%
\frac{3}{2}}$ phase. According to the measurement, $\frac{1}{T_{1}}\propto
T^{0.5}$ above $T_{c}$. Comparing this behavior to (\ref{T1Theory}), we see
that it corresponds to $K_{+,c}\simeq 1.5$. This is consistent with dominant
superconducting correlations.

The value of $K_{+,c}$ extracted from the NQR data is substantially larger
than the one extracted from our DMRG calculation. Clearly, this is a
significant discrepancy. However, it is well known that $K_{+,c}$ is a
non-universal exponent, and in this case it depends strongly on
microscopic details of the problem.
Since the detailed aspects of the microscopic model are certainly not
``realistic,'' we do not feel that any microscopic calculation can
be expected to predict the experimental value of $K_{+,c}$ reliably. A
better strategy is thus to extract $K_{+,c}$ directly from experiments. As
we saw, in the superconducting sample this yields a value of $K_{+,c}$ that
corresponds to dominant superconducting fluctuations.

The NQR measurement was done on with a sample with an oxygen content $%
\delta=0.5$ which is close to the optimal value for superconductivity. We
are not aware of any similar NQR data on 
samples with a different O content and $T_c$.
Were such data obtained, we would predict a smaller power should govern the $%
T$ dependence of 
$1/T_1(T)$, corresponding to a lower $K_{+,c}$. 

\bigskip

\section{Conclusions}

To summarize, we have considered the possibility of zigzag chain-driven
superconductivity in Pr$_2$Ba$_4$Cu$_7$O$_{15-\delta}$. Assuming that the
chains are weakly coupled, this implies that the single zigzag chain must
have a large superconducting susceptibility. This, in conjunction with the
fact that Cu NQR experiments do not show any spin gap above $T_c$, raises
the possibility that the zigzag chain is in a phase in which some, but not
all, of the modes of the two-component electron gas are gapped. We have
shown evidence for such a phase using a microscopic copper-oxygen model for
the zigzag chain.

The gapping of some of the relative spin and charge modes enhances
superconducting (as well as other) fluctuations at low temperatures. The
pairing operator is composed of one hole from each chain with zero center of
mass momentum. Since there is no spin gap, triplet and singlet
superconducting fluctuations are equally enhanced.


\textit{Acknowledgements}. 
We thank S. Sasaki and Y. Yamada for discussions and for sharing their data
with us before publication. S. Moukouri and S. R. White are acknowledged for
their help with setting up the DMRG code. Discussions with E. Altman, O.
Vafek and C. Wu are gratefully acknowledged. We thank A. M. Tsvelik for
critically reading this manuscript. E.B. also thanks the hospitality of the
Weizmann institute, were part of this work was done. This work was supported
in part by NSF grant \# DMR -551196 at Stanford.

\appendix

\section{The C$_1$S$_\frac{3}{2}$ phase}

\label{app:c_1s_1.5}

We would like to describe a phase of the two-component electron gas with a
gap in the relative spin and charge sectors, but no gap in the total spin
and charge sectors. This is the phase that seems to emerge from the DMRG\
results. (See Fig. \ref{fig:DMRG_gaps}.) Starting from the Hamiltonian $%
H=H_{LL}+H_{int}$, where $H_{LL}$ is a Luttinger liquid intra-chain
Hamiltonian (\ref{H_LL}) and $H_{int}$ is the most relevant part of a
generic SU$\left( 2\right) $ invariant inter-chain interaction (Eq. \ref%
{H_int_bos}), we see that in order for the spin gap to vanish we must have $%
g_{2}=0$ in (\ref{H_int_bos})). The remaining term is then proportional to $%
\hat {\mathcal{O}}_{s1}=\cos \left( \sqrt{4\pi }\varphi _{-,s}\right) +\cos
\left( \sqrt{4\pi }\theta _{-,s}\right) $. The total scaling dimension of
this term with respect to the Luttinger liquid fixed point is $1-K_{-,c}$,
with $K_{-,c}<1$, so it will grow under an RG\ transformation. At some
scale, $g_{1}$ becomes of the order of unity, and we can replace $\sin
\left( \sqrt{4\pi }\varphi _{-,c}\right) $ and $\cos \left( \sqrt{4\pi }%
\varphi _{-,c}\right) $ by their mean value (since the field $\varphi _{-,c}$
will be pinned to the minimum of the potential). However, since the
interaction contains the cosines of the conjugate fields $\varphi _{-,s}$
and $\theta_{-,s}$ with an equal weights, both fields fluctuate strongly at
the fixed point, and neither is pinned. We are faced with the problem of how
to describe the low-energy properties of the resulting phase.

A remarkably elegant solution for this problem is given in a paper by
Shelton \emph{et al} [\onlinecite{SheltonTsvelik}]. They considered the
problem of two weakly coupled spin $\frac{1}{2}$ chains, but the results are
readily generalized to our case by neglecting the fluctuations in the $%
\varphi _{-,c}$ field, replacing it by its mean value. The Hamiltonian (\ref%
{H_boson}) can then be solved by re-fermionizing of the fields $\varphi
_{\pm ,s}$. We introduce two Dirac fermions, $\psi $ and $\chi $, as
follows:
\begin{eqnarray}
\psi _{R,L} &=&\frac{1}{\sqrt{2\pi a}}e^{i\sqrt{\pi }\left( \theta _{+,s}\pm
\varphi _{+,s}\right) }  \nonumber \\
\chi _{R,L} &=&\frac{1}{\sqrt{2\pi a}}e^{i\sqrt{\pi }\left( \theta _{-,s}\pm
\varphi _{-,s}\right) }
\end{eqnarray}
where $R$ and $L$ denote right or left moving fields. The Hamiltonian for $%
\psi $ is then just a free Dirac Hamiltonian:
\begin{equation}
H_{+}=u_{s}\int dx\left( \psi _{R}^{\dagger }\frac{1}{i}\frac{d}{dx}\psi
_{R}-\psi _{L}^{\dagger }\frac{1}{i}\frac{d}{dx}\psi _{L}\right)
\end{equation}
The interaction Hamiltonian $H_{int}$ (\ref{H_int}) becomes quadratic in $%
\chi $, giving
\begin{eqnarray}
H_{-} &=&u_{s}\int dx\left( \chi _{R}^{\dagger }\frac{1}{i}\frac{d}{dx}\chi
_{R}-\chi _{L}^{\dagger }\frac{1}{i}\frac{d}{dx}\chi _{L}\right)  \nonumber
\\
&+&\frac{i M}{4}\int dx\left( \chi _{R}^{\dagger }\chi _{L}+\chi
_{L}^{\dagger }\chi _{R}^{\dagger }-\text{H.c.}\right)  \label{H_minus}
\end{eqnarray}
Here $M$ depends on $g_{1}$ and the average of the $\varphi _{-,c}$ part in (%
\ref{H_int}). (\ref{H_minus}) is most conveniently diagonalized by writing $%
\chi $ in terms of two Majorana fermions. Adopting the notation of [%
\onlinecite{SheltonTsvelik}], we write
\begin{eqnarray}
\xi _{\nu } &=&\frac{\chi _{\nu }+\chi _{\nu }^{\dagger }}{\sqrt{2}}
\nonumber \\
\rho _{\nu } &=&\frac{\chi _{\nu }-\chi _{\nu }^{\dagger }}{i\sqrt{2}}
\end{eqnarray}
with $\nu =R,L$. Plugging this into (\ref{H_minus}), we get:
\begin{equation}
H_{-}=H_{0}\left[ \xi \right] +H_{M}\left[ \rho \right]
\end{equation}
\begin{equation}
H_{0}\left[ \xi \right] =\frac{u_{s}}{2}\int dx\left( \xi _{R}\frac{1}{i}%
\frac{d}{dx}\xi _{R}-\xi _{L}\frac{1}{i}\frac{d}{dx}\xi _{L}\right)
\end{equation}
\begin{eqnarray}
H_{M}\left[ \rho \right] &=&\frac{u_{s}}{2}\int dx\left( \rho _{R}\frac{1}{i}%
\frac{d}{dx}\rho _{R}-\rho _{L}\frac{1}{i}\frac{d}{dx}\rho _{L}\right)
\nonumber \\
&&+\frac{i M}{2}\int dx\left( \rho _{R}\rho _{L}-\rho _{L}\rho _{R}\right)
\end{eqnarray}
The field $\xi $ is thus massless, while $\rho $ is massive, with a mass $M$%
. In that sense, ``half" of the field $\varphi _{-,s}$ is gapped. The total
spin sector, on the other hand, is completely massless. We therefore denote
this phase as a C$_{1}$S$_{\frac{3}{2}}$ phase.

Setting $g_{2}\neq 0$ in (\ref{H_int}) will gap out also the total spin
mode, giving a C$_{1}$S$_{0}$ phase. Moreover, adding less relevant inter-
or intra-chain operators to (\ref{H_int}) would generate the $g_{2}$ term,
even if it is not present in the bare Hamiltonian. However, the DMRG\
calculation presented in section \ref{sec:DMRG} indicates that the spin gap $%
\Delta _{s}$ is much smaller than $\Delta _{\delta N}$, the gap to
transferring one hole from one chain to the other, which is related to a gap
in the relative spin/charge modes. Therefore, at intermediate energies (or
temperatures) between $\Delta _{s}$ and $\Delta _{\delta N}$, the physics is
expected to be dominated by the unstable C$_{1}$S$_{\frac{3}{2}}$ fixed
point.

Next, let us find the long-distance behavior of the correlation functions of
physical operators in the C$_{1}$S$_{\frac{3}{2}}$ phase. For concreteness,
we will focus on the pair field operator $\Delta _{SC}=\psi _{1R,\uparrow
}\psi _{2L,\downarrow }$. This operator creates a pair of holes in the two
chains with opposite momenta and spins. In bosonized form, this operator
becomes
\begin{equation}
\Delta _{SC}\sim e^{i\sqrt{\pi }\left( \theta _{+c}+\theta _{-,s}+\varphi
_{-,c}+\varphi _{+,s}\right) }
\end{equation}
$\theta _{+c}$ is a free bosonic field, whose long-range correlations are $%
\left\langle e^{i\sqrt{\pi }\theta _{+,c}\left( x\right) }e^{i\sqrt{\pi }%
\theta _{+,c}\left( 0\right) }\right\rangle \sim x^{-\frac{1}{2K_{+,c}}}$.
Similarly, $\left\langle e^{i\sqrt{\pi }\varphi _{+,s}\left( x\right) }e^{i%
\sqrt{\pi }\varphi _{+,s}\left( 0\right) }\right\rangle \sim x^{-\frac{1}{2}%
} $, since $\varphi _{+,s}$ is a free field with $K_{+,s}=1$, as dictated by
the SU$\left( 2\right) $ symmetry. This correlation function is expected to
have logarithmic corrections due to marginal operators, which we have
neglected. The $\varphi _{-,c}$ field is massive, so at long distances we
can replace $e^{i\sqrt{\pi }\varphi _{-,c}}$ by its expectation value. The
treatment of $e^{i\sqrt{\pi }\theta _{-,s}}$ is more subtle, since as we
have mentioned, this field is ``half gapped" in the C$_{1}$S$_{\frac{3}{2}}$
phase. It is nevertheless possible to calculate its correlations, following
a method used in [\onlinecite{SheltonTsvelik}]. We describe this method
briefly here. The relative spin sector, which is described as two
independent Majorana theories, can be further mapped onto two 2d Ising
models. Since one of the Majorana fields is massless and the other is
massive, one of the corresponding Ising fields is at criticality and the
other is away from criticality. It can be shown that the four operators $%
\sin \left( \sqrt{\pi }\theta _{-,s}\right) $, $\cos \left( \sqrt{\pi }%
\theta _{-,s}\right) $, $\sin \left( \sqrt{\pi }\varphi _{-,s}\right) $, $%
\cos \left( \sqrt{\pi }\varphi _{-,s}\right) $ have the following
correspondance to the order and disorder operators of the two Ising models%
\cite{SheltonTsvelik,ShankarBosonization}:
\begin{eqnarray}
\cos \sqrt{\pi }\varphi _{-,s} &=&\mu _{1}\mu _{2}\text{, \ \ \ }\sin \sqrt{%
\pi }\varphi _{-,s}=\sigma _{1}\sigma _{2}  \nonumber \\
\cos \sqrt{\pi }\theta _{-,s} &=&\mu _{1}\sigma _{2}\text{, \ \ \ }\sin
\sqrt{\pi }\theta _{-,s}=\sigma _{1}\mu _{2}  \label{operator_correspondance}
\end{eqnarray}
Here $\sigma _{1}$, $\mu _{1}$, $\sigma _{2}$, $\mu _{2}$ are the
order/disorder operators of the two Ising models labelled 1 and 2. Note that
the scaling dimension of all the operators in the left hand side of (\ref%
{operator_correspondance}) at criticality is $\frac{1}{4}$, which is
consistent with the fact that the dimension of the Ising operators at
criticality is $\frac{1}{8}$. Carrying out the correspondence to the Ising
model carefully, one finds that the Ising model labelled by 1 is in its
disordered phase, so $\left\langle \mu _{1}\right\rangle \neq 0$, $%
\left\langle \sigma _{1}\right\rangle =0$. The other Ising model labelled by
2 is critical, and therefore $\left\langle \sigma _{2}\left( x\right) \sigma
_{2}\left( 0\right) \right\rangle \sim x^{-\frac{1}{4}}$, and similarly for $%
\mu _{2}$. Thus at long distances, $\left\langle e^{i\sqrt{\pi }\theta
_{-,s}\left( x\right) }e^{i\sqrt{\pi }\theta _{-,s}\left( 0\right)
}\right\rangle \sim x^{-\frac{1}{4}}$. The exponent of this correlation
function is just half of the exponent that one gets for a free $\theta
_{-,s} $ field. This can be understood as a consequence of the fact that
``half" of the $-,s$ mode is gapped. Therefore, correlation function of $%
\Delta _{SC}$ behaves as
\begin{equation}
\left\langle \Delta _{SC}^\dagger \left( x\right) \Delta _{SC}\left(
0\right) \right\rangle \sim \frac{1}{x^{\frac{1}{2K_{+,c}}+\frac{3}{4}}}
\end{equation}
which gives a superconducting susceptibility that behaves as $\chi
_{SC}\left( T\right) \sim T^{\frac{1}{2K_{+,c}}-\frac{5}{4}}$. The
correlation functions of other physical operators (CDW, SDW etc.) can be
calculated in a similar manner.

It is interesting to note that the Hamiltonian (\ref{H_int_bos}) with $g_2=0$
is equivalent to the integrable super-symmetric super-Sine-Gordon model\cite%
{TsvelikSSGmodel}.

\end{document}